\documentclass[a4paper,11pt]{article}
\usepackage{pos}
\usepackage{slashed}

\usepackage{graphicx}  
\usepackage{dcolumn}   
\usepackage{bm}        
\usepackage{physics}
\usepackage{tikz}
\usepackage{mathtools}
\usepackage{amsmath} 

\newcommand{\QQbar}{{Q\bar Q}}

\newcommand{\mcA}{{\mathcal A}}
\newcommand{\mcAbar}{{\overline\mcA}}

\newcommand{\lp}{\left}
\newcommand{\rp}{\right}

\FullConference{The 40th International Symposium on Lattice Field Theory (Lattice 2023)\\
July 31st - August 4th, 2023\\
Fermi National Accelerator Laboratory\\}

\title{Entanglement Entropy due to the Presence of Static Quarks}

\author*[a]{Rocco Amorosso}
\author[a]{Sergey Syritsyn}

\affiliation[a]{Department of Physics and Astronomy, Stony Brook University, 
  Stony Brook, New York 11794, USA}
\date{\today}%

\emailAdd{rocco.amorosso@stonybrook.edu}
\emailAdd{sergey.syritsyn@stonybrook.edu}

\abstract{
We study the entanglement of gluon fields in presence of a static $\QQbar$ pair in quenched QCD.
Using the replica method, we investigate the $q=2$ Renyi entropy of the entanglement of gluon fields
inside and in the vicinity of the confining QCD string between the quark and the antiquark.
We find that there is excess entropy of gluon entanglement compared to vacuum fluctuations.
This excess of entanglement entropy is associated with the gluon flux tube,
and we find that it has a finite non-zero value in the continuum.
We investigate the dependence of gluon entanglement on the geometry of longitudinal and transverse
partitioning of the flux tube.
Our preliminary results suggest scaling of the entanglement entropy with the area of the
boundary overlapping with the flux tube.
}

\FullConference{%
 The 40th International Symposium on Lattice Field Theory, LATTICE2023
  July 31 - Aug 4, 2021
}

\begin{document}
\maketitle

\section{Introduction}
Dynamics of the QCD string play a large role in many phenomena, from hadronization in heavy ion
collisions to deep inelastic scattering.
In the Lund model, an expanding QCD string is assumed to break into fragments that hadronize
independently, 
even though it is a pure spatially extended state of gluons at zero
temperature\cite{Ferreres-Sole:2018vgo}.
At the same time, thermal-like behavior in hadronization is difficult to explain due to a lack of
hadron rescattering~\cite{Fischer:2016zzs}.
On the other hand, in toy models exhibiting confinement, the reduced density matrix of an expanding
string has been shown to take a thermal form, helping to explain the success of thermal models in
$e^+e^-$ collisions\cite{Berges:2018cny}.
Given the complex dynamics of the QCD string, it is of much interest to study the QCD string on the lattice
from the perspective of quantum entanglement.

Quantum entanglement has offered a window into better understanding phase transitions and the
emergence of colorless confined states through studying the entropic C-function of entanglement
of a bi-partitioned Yang-Mills vacuum~\cite{Buividovich:2008kq,Buividovich:2008gq,Itou:2015cyu,Rabenstein:2018bri}.
As the size $l$ of the sub-region $A$ was varied, a transition from $O(N_c^2)$ to $O(1)$ degrees of
freedom was observed, representing a transition from colorful quarks and gluons at small distances to
colorless hadrons and glueballs at larger
distances~\cite{Buividovich:2008kq,Buividovich:2008gq,Itou:2015cyu,Rabenstein:2018bri}.

It is plausible that the thermal spectrum in hadronization is a result of quantum entanglement within an
approximately pure quantum state of an expanding string between a quark and
antiquark~\cite{Berges:2017zws}.
In this work, we perform a novel study of the entanglement entropy of gluon fields in and around a QCD
string between a heavy quark-antiquark pair.
The entanglement entropy is calculated using the temporal replica method similar to
Refs.~\cite{Buividovich:2008kq,Rabenstein:2018bri} but applied to the ground state of a QCD flux
tube.
In particular, we explore whether the excess entropy due to the presence of the string is UV-finite.

\section{Entanglement entropy in a QCD string on a lattice}
Entanglement entropy in field theory is typically defined for quantum fields contained in
a region $\mcA$ and its complement $\mcAbar$.
Following Refs.~\cite{Buividovich:2008kq,Rabenstein:2018bri}, we divide lattice gauge links into
regions $\mcA$ and $\mcAbar$ along the gauge links; 
the links on the boundary are assigned to $\mcA$.
Decomposing the complete Hilbert space of non-abelian gauge theories
$\mathcal{H} = \mathcal{H}_\mcA \otimes \mathcal{H}_\mcAbar$ in this way has subtleties discussed in
Ref.\cite{Buividovich:2008kq} where it is argued that such partition is maximally gauge-invariant.

Entanglement entropy is computed from the reduced density matrix
\begin{equation}
\label{eqn:rho_reduced}
\hat{\rho}_\mcA=\Tr_{\mcAbar}\hat{\rho}\,,
\end{equation}
where $\hat{\rho}$ is the full density matrix of the system and $\Tr_{\mcAbar}$ represents the partial trace over quantum fields in region $\mcAbar$.
The von Neumann entanglement entropy is then
\begin{equation}
   S^\text{EE}=-\Tr \lp(\hat{\rho}_\mcA \log \hat{\rho}_\mcA\rp),
\end{equation}
as it most closely resembles the Boltzmann statistical entropy.
However, explicitly calculating the density matrix or the von Neumann entanglement entropy on the
lattice is in our context intractable, so instead we employ the Renyi entropy of order $q$:
\begin{equation}
    S^{(q)}=\frac{1}{1-q}\log\lp(\Tr\hat{\rho_\mcA}^q\rp) .
\end{equation}
This instead requires $\Tr\hat{\rho}_\mcA^q$, which can be realized on a
lattice using replicas and non-trivial boundary conditions.
One can obtain the von Neumann entropy in the limit 
\begin{equation}
    S^\text{EE}=\lim \limits_{q\to 1} S^{(q)}\,,
\end{equation}
which requires analytic continuation in $q$. 
Since only integer $q$ are possible on a lattice, one would have to rely on interpolation (e.g., using Pad\'e
approximant) to accomplish this.

In field theory, entanglement entropy is typically UV-divergent. 
However, the QCD string is a physical object with finite energy density and profile width, 
so it can be expected to have finite contribution to the entanglement entropy when it is 
partitioned into regions $\mcA$ and $\mcAbar$.
To investigate this, we compute the difference between Renyi entropies of gluon entanglement 
in the presence of a pair of static color sources $\QQbar$ and in vacuum:
\begin{equation}
\label{eqn:RenyiDiff}
    \tilde{S}^{(q)}_{\vert Q \bar{Q}} = S^{(q)}_{\vert Q \bar{Q}}-S^{(q)} .
\end{equation}

\begin{figure}[ht!]
  \centering
  \includegraphics[width=.4\textwidth]{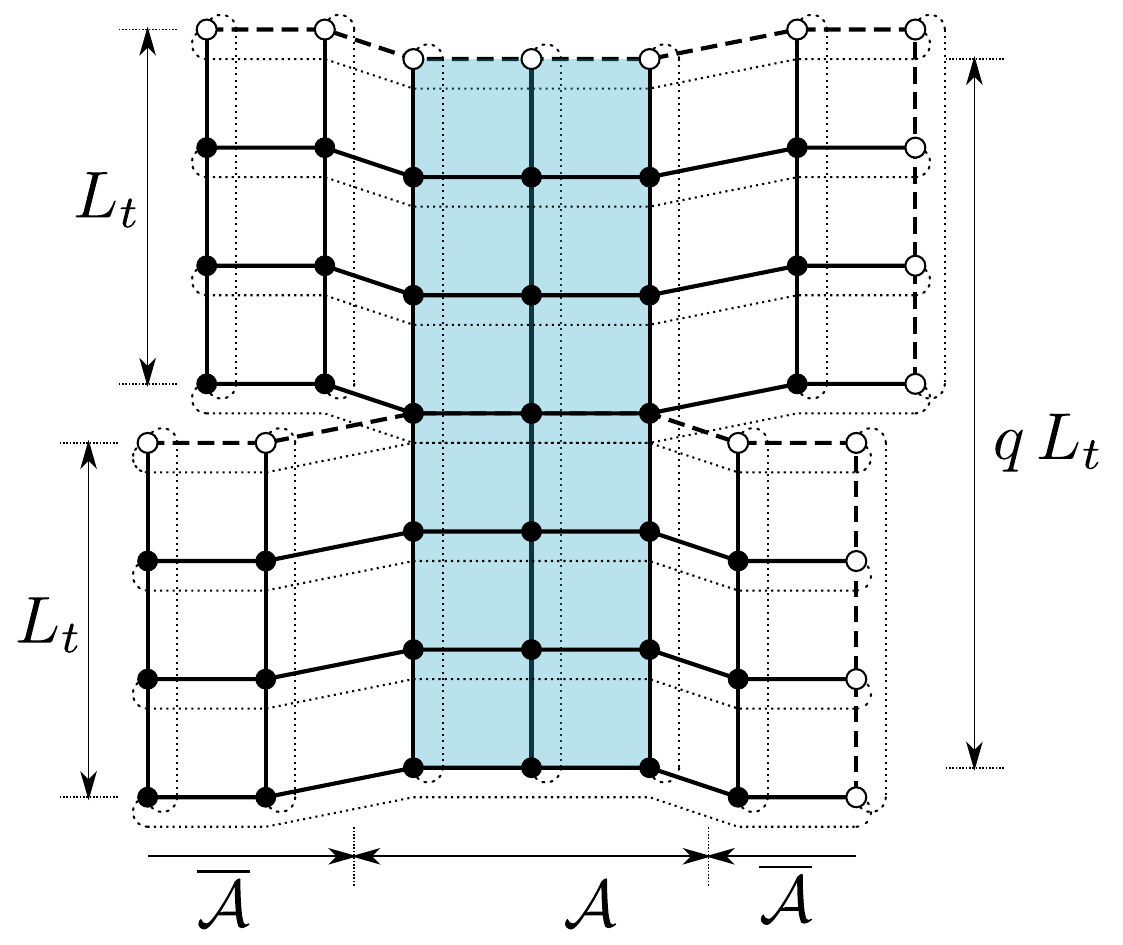}
  \caption{\label{fig:pants}
Illustration of the geometry of $q=2$ replicas of an $L_x\times L_\tau=6\times3$ lattice. The spatial dimension $x$ is
horizontal and the time dimension $\tau$ vertical.
Region $\mcA$ is shaded blue.
Open points on the edge are periodic wrap-around images of the sites to which they are 
connected by dashed lines.
The boundary conditions in the $x$ direction are periodic.
}
\end{figure}

For computing Renyi entanglement entropy on a lattice, we follow the same procedure as
Ref.~\cite{Itou:2015cyu,Rabenstein:2018bri} to generate powers of the partially-traced density matrix
$(\hat{\rho}_\mcA)^q$ of gluon fields.
As described in Ref.\cite{Calabrese:2004eu}, $q$ independent replicas of the lattice, each of temporal
dimension $L_t=\beta=T^{-1}$, are stacked in temporal direction.
In region $\mcAbar$, the time boundary conditions for the links are 
$\phi_{\tau=\beta}^{(r)}=\phi_{\tau=0}^{(r)}$ for each replica $r=1\ldots q$,
which correspond to traces over links in $\mcAbar$ only.
In region $\mcA$ however, the link variables are identified 
for consecutive replicas $r$ and $r+1$ as $\phi_{\tau=\beta}^{(r)}=\phi_{\tau=0}^{(r+1)}$, 
and also $\phi_{\tau=\beta}^{(q)}=\phi_{\tau=0}^{(1)}$ for the overall trace of $\rho_\mcA^q$.
This geometry is illustrated in Fig.~\ref{fig:pants} for $q=2$.

We place the quark and the antiquark in the complement region $\mcAbar$,
and their color degrees of freedom are traced according to Eq.~(\ref{eqn:rho_reduced}),
resulting in Polyakov loops.
Therefore, the excess of entanglement entropy due to the QCD string created by a $\QQbar$
pair~(\ref{eqn:RenyiDiff}) can be calculated directly from the ratio of Polyakov loop correlators.
\begin{equation}
\label{eqn:RenyiDiffPolyakov}
\tilde{S}^{(q)}_{\vert Q \bar{Q}} 
= -\frac{1}{q-1} \log
  \frac{\langle \prod\limits_{r=1}^{q}\Tr P_0^{(r)}\Tr P_{\vec{x}}^{(r)\dag}\rangle}
       {\big[\langle\Tr P_0\Tr P_{\vec{x}}^\dag\rangle\big]^q},
\end{equation}
where $\Tr P_{\vec{x}}^{(r)}$ represents the Polyakov loop at location $\vec{x}$ in replica $r$,
and $\Tr P_{\vec{x}}$ represents the usual Polyakov loop on the standard periodic (one-replica) lattice.

Calculations in this work are performed at half of the critical temperature,
at which the string tension is within 1\% of that at zero temperature~\cite{Cardoso:2011hh},
and we assume that the string is sufficiently close to the ground state.
Calculation at a lower temperature would require substantially more statistics due to increased noise
in the Polyakov loops.
We use the standard Wilson plaquette $SU(3)$ Yang-Mills action, performing sweeps with alternating
Kennedy-Pendleton heatbath updates \cite{Kennedy:1985nu} and over-relaxation updates
\cite{Brown:1987rra}, as well as the multilevel algorithm~\cite{Luscher:2001up} 
for computing the Polyakov loop correlators. 
The scale $a$ is set by the string tension $\sigma=(440\,\text{MeV})^2$ following
Ref.~\cite{Cardoso:2011hh}.

\section{Numerical results}

\begin{figure}
  \centering
  \includegraphics[width=.25\textwidth]{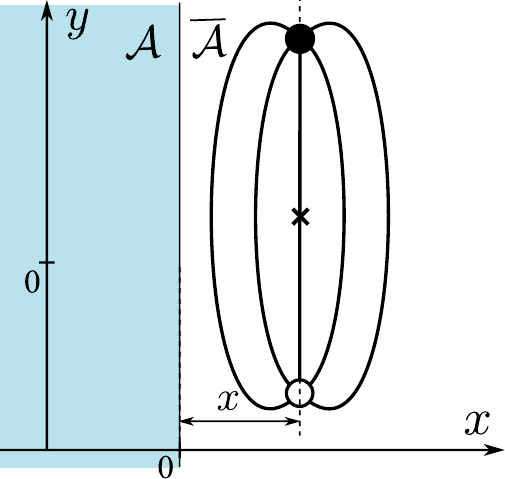}~
  \hspace{.10\textwidth}~
  \includegraphics[width=.25\textwidth]{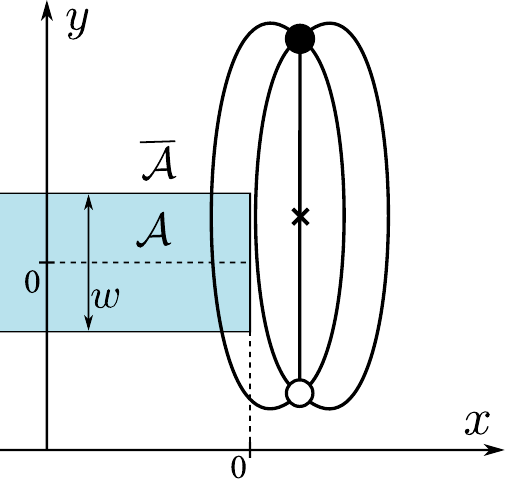}\\
  \caption{\label{fig:crudehalfslab}
    In this and subsequent figures, point $(x,y)$ corresponds to the middle of the flux tube 
    connecting the $\QQbar$ pair along the $\hat y$-axis.
    (Left) The $\QQbar$ string is parallel to the boundary of $\mcA$ (``slab'' $-w\le x\le0$), 
    and sliced into partitions residing in $\mcA$ and $\mcAbar$.
    (Right) The $\QQbar$ string is perpendicular to $\mcA$ 
    (``half-slab'' $-\infty<x\le0$ and $|y|\le\frac w2$).
    The string is wholly in region $\mcAbar$ at $x\to\infty$, 
    and cross-cut into segments at $x\to(-\infty)$.
    At $x\sim0$, the flux tube is partially cross-cut.
}
\end{figure}

\begin{figure}
    \centering
    \includegraphics[width=.48\textwidth]{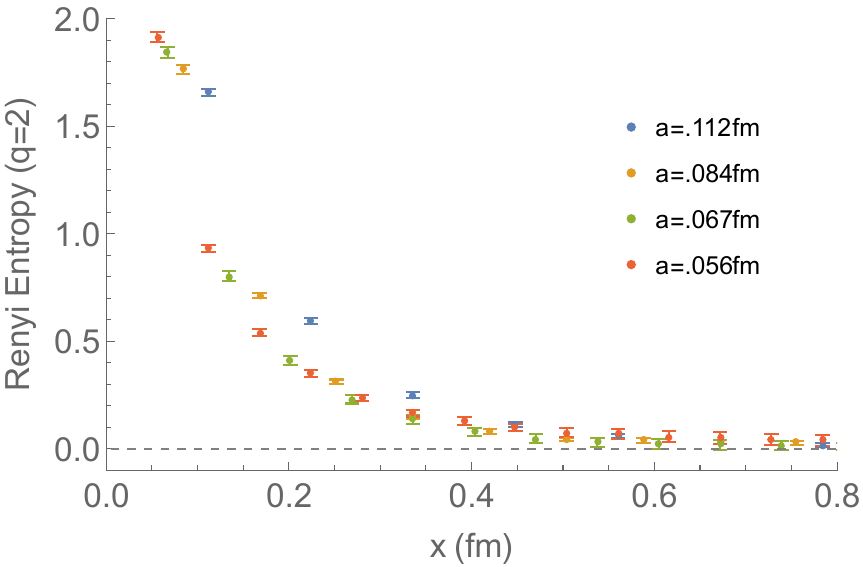}
    \caption{
      The excess entanglement entropy of a flux tube sliced parallel to $\QQbar$ separation
      $d_{\QQbar}=0.336$ fm (see Fig.~\ref{fig:crudehalfslab}, left).
      Dependence on the distance to the $\mcA/\mcAbar$ boundary compared at different lattice spacings.
}
    \label{fig:perpconfined}
\end{figure}

We first study the entanglement of the string ``sliced'' into transverse partitions.
We make region $\mcA$ an infinite slab of width $w=0.336$ fm parallel to the flux tube connecting
the $\QQbar$ pair (see Fig.~\ref{fig:crudehalfslab}, left).
The $\QQbar$ pair is separated by $d_\QQbar=0.336$ fm.
In Figure~\ref{fig:perpconfined}, we show the excess entanglement entropy~(\ref{eqn:RenyiDiff})
$\tilde S^{(2)}_{|QQ}$ as a function of $x$, the distance from the $\QQbar$
pair to the $\mcA/\mcAbar$ boundary.
Comparing the results with lattice spacings $a=0.056\ldots0.112$ fm, we observe that 
$\tilde S^{(2)}_{|\QQbar}(x)$ has no UV divergence and is finite in the continuum limit.
Also, except for apparent discretization artifacts at the coarsest lattice spacing $a=0.112$ fm, the
data points appear to line up as the lattice spacing is decreased.
This indicates convergence to a universal profile $\tilde S^{(2)}_{|QQ}(x)$, which is nonzero within
the expected width of the flux tube~\cite{Bicudo:2017uyy}.
Performing a continuum limit extrapolation is complicated by different discrete values of $x$ at different values of $a$.


\begin{figure}[ht!]
  \centering
  \includegraphics[width=.48\textwidth]{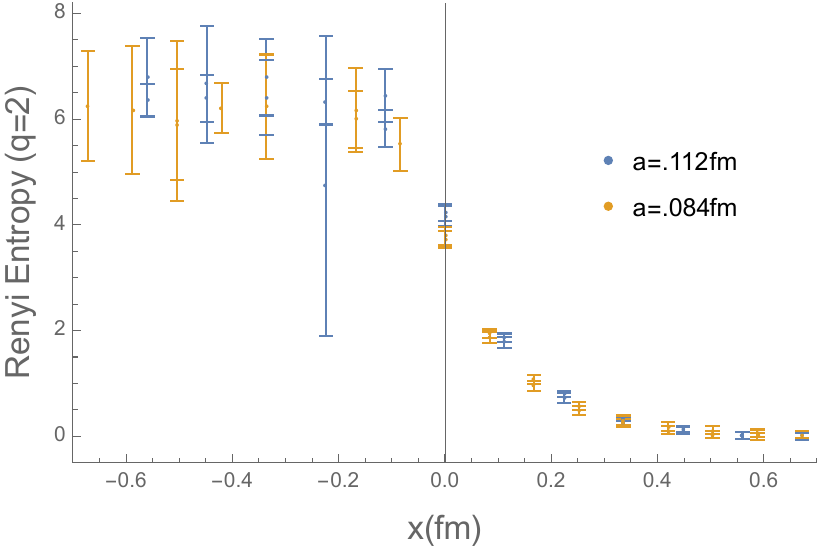}
  \caption{\label{fig:sEE_perp_ascaling}
    The excess entanglement entropy of the flux tube (partially) cross-cut 
    by region $\mcA$, see Fig.~\ref{fig:crudehalfslab}(right) for explanation.
    Comparison of results obtained with two lattice spacings and fixed $\QQbar$ distance $d_\QQbar=0.672$ fm 
    and region $\mcA$ width $w=0.336$ fm. 
  }
\end{figure}

To study the entanglement of gluons in a flux tube ``cross-cut'' into length-wise segments, we make the region
$\mcA$ in the shape of a slab with width $w$ placed between the quark and antiquark.
We have found that decreasing the length of the slab in the transverse direction to span only half
of the lattice (i.e., $\mcA=w\times \frac12{L_\sigma} \times L_\sigma$, $L_\sigma$ is the spatial
extent of the lattice) leads to a dramatic
reduction of statistical fluctuation in the Polyakov loop correlators~(\ref{eqn:RenyiDiffPolyakov}).
At the same time, this has the benefit of probing the entanglement entropy as the flux tube is
gradually ``cross-cut'' by $\mcA$ (see Fig.~\ref{fig:crudehalfslab}, right), and allows us to examine both the
longitudinal and transverse features of the string's entropy profile.
We refer to this setup as the ``half-slab'' geometry.
The excess entanglement entropy of a completely cross-cut flux tube $x\to-\infty$ shows little dependence
on the lattice spacing (see Fig.~\ref{fig:sEE_perp_ascaling}), indicating that it is also UV-finite.

\begin{figure}
    \centering
    \includegraphics[width=.48\textwidth]{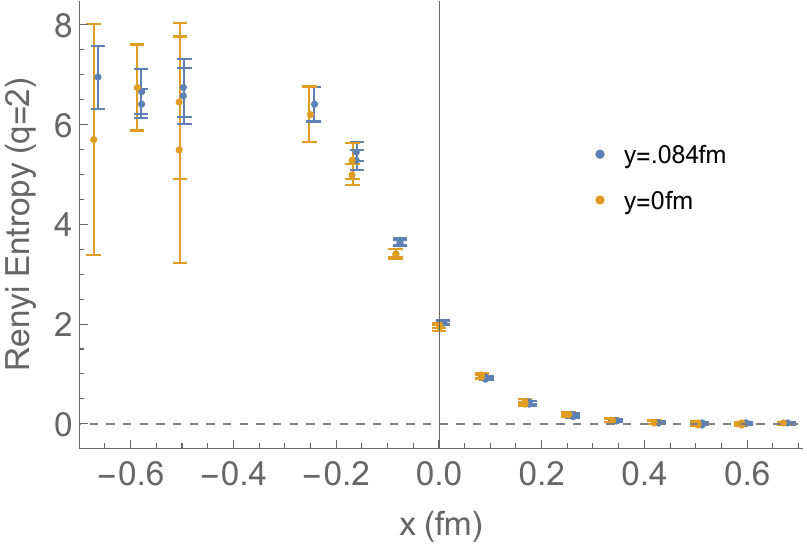}~
    \hspace{.02\textwidth}~
    \includegraphics[width=.48\textwidth]{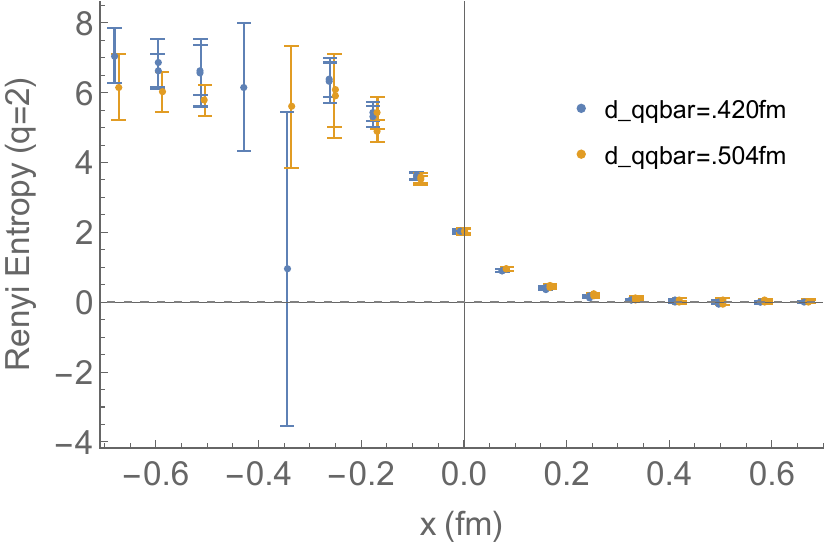}\\
    \caption{\label{entropylocationdependence}
      The excess entanglement entropy of the flux tube (partially) cross-cut 
      by region $\mcA$ with $w=a=0.084$ fm, see Fig.~\ref{fig:crudehalfslab} for explanation.
      (Left) dependence on the position $y$ relative to the center of region $\mcA$
      The $\QQbar$ distance is $d_\QQbar = 5a = 0.420$ fm.
      (Right) dependence on the distance between $Q$ and $\bar Q$. 
      The data points are offset horizontally for clarity.
}
\end{figure}

Next, we examine the dependence of the entanglement on the length of the flux tube and the length of its segments that reside in $\mcA$ and $\mcAbar$.
In Figure~\ref{entropylocationdependence}(left), we compare $\tilde S_{|\QQbar}$ as the region $\mcA$ 
is shifted longitudinally relative to $Q$ and $\bar Q$.
This results in varying the lengths of the string segments in $\mcAbar$ (attached to $Q$ and $\bar Q$),
while the length of the middle segment in $\mcA$ is constant.
We observe no statistically significant dependence on $y$ at any values of $x$, including $x\sim0$, 
when the flux tube is partitioned in both $x$ and $y$ directions.
When the string is completely cut ($x\to-\infty$), 
we also do not find any dependence of $\tilde S_{|\QQbar}$ on the length of the string 
$\QQbar$ separation $d_\QQbar=0.420\ldots0.504$ fm (see Fig.~\ref{entropylocationdependence}, right).

\begin{figure}
  \centering
  \includegraphics[width=.48\textwidth]{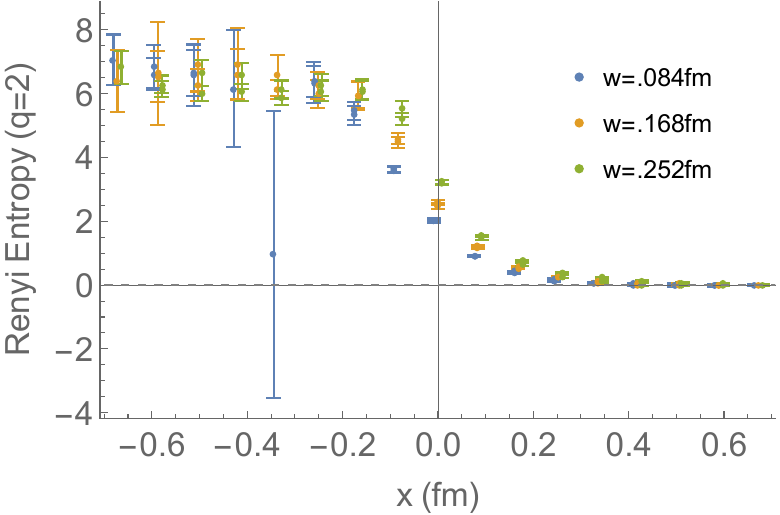}~
  \hspace{.02\textwidth}~
  \includegraphics[width=.48\textwidth]{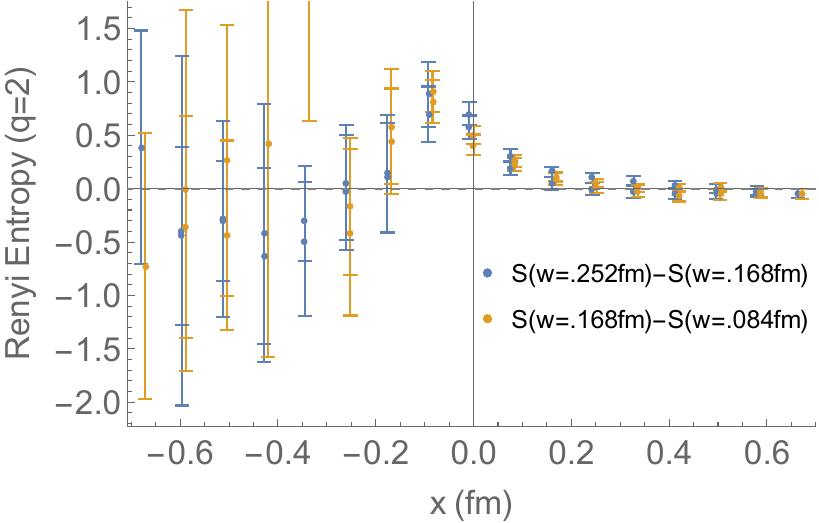}\\
  \caption{\label{entropyregionwidthdependence}
    The excess entanglement entropy of the flux tube (partially) cross-cut 
    by region $\mcA$ of varying width $w$, see Fig.~\ref{fig:crudehalfslab} for explanation.
    (Left) dependence on the width $w$ of region $\mcA$, and
    (right) difference of entanglement entropies between $w$ and $w+a$.
    The $\QQbar$ distance is $d_\QQbar = 5a = 0.420$ fm.
    The data points are offset horizontally for clarity.
}
\end{figure}

Finally, we examine the dependence of $S_\QQbar^{(2)}$  on the width $w$ of region $\mcA$.
When the flux tube is completely cut ($x\to-\infty$), we find no observable dependence as 
$w$ is varied in range $\approx0.08\ldots0.25$ fm (Fig.~\ref{entropyregionwidthdependence}, left).
However, there is a prominent dependence on the width when the flux tube is partially cut ($x\sim0$).
To study this effect, we calculate the finite difference $\tilde S_{|\QQbar}\Big|_{w}^{w+a}$
and find that $\tilde S^{(2)}_{|\QQbar}$ increases roughly linearly with $w$ 
(Fig.~\ref{entropyregionwidthdependence}, right).
This can be interpreted as an approximate area law: 
\emph{the excess entanglement entropy of the fields comprising the string is proportional to the
area of the boundary between $\mcA$ and $\mcAbar$ overlapping with the QCD flux tube}.

\section{Discussion}

In this work, we have computed the entropy of entanglement between parts of a quenched QCD string
between two static quarks on a lattice.
By subtracting the vacuum contribution, we manage to obtain UV-finite values for the entropy.
We have explored the entanglement of fragments of a color flux tube ``sliced'' parallel and
``cross-cut'' perpendicular to its orientation.
In the transverse direction, the entanglement entropy decays over the distance consistent with 
its width.
In the longitudinal direction, we have found that the entanglement of segments of a color flux tube
depends very weakly on the $\QQbar$ separation, the position of the central segment, and its length.
However, for a partially cross-cut string, the entanglement entropy grows roughly linearly with the
area of the $\mcA/\mcAbar$ boundary overlapping with the flux tube.
Some of these observations have been made at only one lattice spacing $a=0.084$ fm, and a thorough
continuum-limit study must be conducted.

Further questions to investigate include the dependence of entanglement entropy on the temperature,
number of colors, and string length up to $d_\QQbar\gtrsim1$ fm.
Ultimately, entanglement in a dynamically breaking QCD string must be investigated,
which is not feasible with dynamical quarks at the moment.
However, a tentative study can be performed in quenched QCD on an adjoint string.

\section*{Acknowledgements}
R.A. and S.S. are supported by NSF supported by the National Science Foundation under CAREER Award
PHY-1847893.
In addition, R.A. was supported in part by the Office of Science, Office of Nuclear Physics,
U.S. Department of Energy under Contract No. DEFG88ER41450 and 
 by the Simons Foundation under Award number 994318 (Simons Collaboration on Confinement and QCD Strings).
The authors thank Stony Brook Research Computing and Cyberinfrastructure and the
Institute for Advanced Computational Science at Stony Brook University for access to the Seawulf HPC
system, which was made possible by grants from the National Science Foundation (awards 1531492 and
2215987) and matching funds from the Empire State Development’s Division of Science, Technology
and Innovation (NYSTAR) program (contract C210148).

\bibliography{entent-lat23proc}
\bibliographystyle{aip_ep}

\end{document}